\documentclass[twocolumn,showpacs,amsmath,amssymb,prb]{revtex4}

\usepackage{amsmath}
\usepackage{amssymb}
\usepackage{graphicx}
\usepackage{dcolumn}
\usepackage{bm}

\def\be{\begin{eqnarray}}
\def\ee{\end{eqnarray}}
\def\ba{\begin{array}}
\def\ea{\end{array}}

\begin{document}
\author{D. Eksi}
\address{Vacational School of Health, Yeni Yuzyil University, Istanbul, 34010, Turkey}
\title{The topological inequivalence of Hall bar versus Corbino geometries in real space}
\author{A. Yildiz Tunali}
\address{Dokuz Eylul University, Physics Department, Faculty of Arts and
Sciences, Tinaztepe Campus, 35160-Buca, Izmir, Turkey}
\author{A. Siddiki}
\address{Paul-Drude-Institut fuer Festkörperelektronik, Leibniz-Institut im Forschungsverbund Berlin e.V., Hausvogteiplatz 5-7, 10117 Berlin, Germany}
\address{Physics Department, Ekendiz Tanay Center for Arts and Science, Ula, Mugla 48650, Turkey}

\begin{abstract}
This work discusses the effect of topology in the frame of direct Coulomb interactions, considering two distinct geometries, namely the Hall bar and the Corbino disc. In the mainstream approaches to the quantized Hall effect, the consequences of interactions are usually underestimated. Here, we investigate the electron number density, potential and current distributions within the screening theory that considers electron-electron interactions. Inclusion of direct Coulomb interaction and realistic boundary conditions result in local metal-like compressible and (Topological) insulator-like incompressible regions. Consequently, we show that the bulk of both geometries in coordinate space is not incompressible throughout the quantized Hall plateau. Furthermore, placing two inner contacts within the Hall bar geometry shows that the quantization is unaffected by changing the genus number in real space. Finally, we propose novel experiments which will enable us to distinguish the topological properties of the two geometries in the configuration space.
\end{abstract}


\maketitle

\section{Introduction}
The comprehensive review by Hassan and Kane~\cite{Kane10:3045} describes the quantum Hall effects~\cite{vKlitzing80:494,FQHE} (QHEs) observed at two-dimensional electron systems (2DES) at low temperatures and subjected to perpendicular high magnetic fields $B$, as a cousin of Topological Insulators (TI). This argument derives its rationale from the gauge invariance principle, which several authors have discussed since the early '80s.~\cite{Laughlin81:5652,Thouless81:3475,Halperin82:2185,Simon83:2167,Kohmoto85:343} Starting with the pioneering work of Laughlins's Gedankenexperiment considering a cylinder where $B$ is perpendicular to the curved 2D surface.~\cite{Laughlin81:5652} The influencing paper follows this work by Halperin,~\cite{Halperin82:2185} providing explicit expressions in describing quantized charge transport considering a Corbino geometry (an annular disc with a hollow region at the centre). The latter work discusses the possibility of localized and extended states, and the localization is focused on in a consequent paper by Thouless.~\cite{Thouless81:3475} Finally, the boundary condition sensitivity of the integer QHE (IQHE) is discussed using a periodic
potential was first discussed by Thouless {\emph{et al.}}.~\cite{Thouless82:405} This seminal
work, known as the TKNN theory, strongly criticize the techniques in calculating the conductance quantization expressions' insensitivity to the boundary conditions, which is also one of the major concerns of the present paper.

The topological character of the two-dimensional electron system (2DES) is described in the later works by Simon~\cite{Simon83:2167} and Kohmoto,~\cite{Kohmoto85:343} which consider Bloch type periodic wavefunctions with translational invariance to preserve the gauge invariance. The wavefunction remains invariant when a phase is attributed depending on $B$, the Aharonov-Bohm phase,~\cite{Aharonov59:485} a particular case of Berry's phase.~\cite{Berry84:45} However, these calculations are performed in the momentum space considering a Genus one (toroidal) topology, while a geometry independent energy gap opens due to a quantizing strong magnetic field, i.e., Landau quantization. 

It is imperative to note that all the above theories perform analytical calculations in the momentum space. The momentum space is dual to real space with the boundary conditions imposed. However, as we will discuss, such boundary conditions this is not true for actual sample geometries, which also have current and voltage contacts.

It is stated that the topological insulators (materials with a bulk insulating gap in the energy-momentum dispersion, not in energy-coordinate space) exhibit quantum-Hall-like behaviour without a magnetic field. Namely, the bulk is topologically insulating (incompressible); meanwhile, the edges are conductive. The topological states are conventional materials conducted in momentum space utilizing methods like ARPES given in Ref~\onlinecite{Kane10:3045} and the references therein. However, to our knowledge, no direct charge transport experiments are performed considering TI, probably because no transport contacts can be realized at the momentum space. 

In contrast to TI, direct transport measurements at real space with two inner contacts at the bulk were proposed by the authors of the present paper,~\cite{Yildiz14:014704} where they discussed the charge transport within the non-linear screening theory of the QHE by computing the capacitances between inner contacts and estimating the Hall resistance corresponding. The approach rationale stems from the strong relationship between compressibility and capacitance via their dependence on the density of states (DOS). The proposed direct transport experiments were performed at high-quality narrow Hall bars~\cite{Kendirlik17:14082} confirming that the bulk of the sample in coordinate space is not incompressible for all $B$ strengths throughout the quantized Hall Plateau.

The interpretation of the IQHE within the screening theory, which also takes into account direct Coulomb interactions, is in contrast with the commonly used Landauer-B\"{u}ttiker-based edge state theory.~\cite{Buettiker86:1761,Buettiker88:9375} In the non-interacting edge-state picture, the equilibrium current flows along with independent and spatially separated 1D edge channels in counter directions at different sides. Note that, here, sides refer to physical left and right sides of the sample, which is just a consequence of the assumption of a translational invariant sample where $+k$, i.e. momentum of the forward-moving electrons, can be directly taken to be right edge of the sample. Namely, one can replace $+k$ with $+x$, and the same is true for the back-moving electrons. This picture is valid only if an equilibrium current is considered with translational invariance. The number of edge-states is determined by the number of states beneath the Fermi energy $E_F$ at each quantized Hall plateau. The quantization is obtained simply by counting the number of these states. For example, at the 2$^{th}$ plateau, there are four edge-states carrying equilibrium current, two in forward-moving direction and two in the back-moving direction. Actual Hall bar samples are neither translational invariant nor periodic once a non-equilibrium current is imposed via source-drain contacts and measured by probe contacts.

In contrast, in the screening theory Gerhardts and his co-workers developed, the imposed non-equilibrium uni-directional current is confined to one incompressible (edge or bulk) state with a respective filling factor during the existence of a quantum Hall plateau.~\cite{Lier94:7757,Oh97:108,Guven03:115327,SiddikiNJP10:113011} Their observation and conclusion are in perfect agreement with the local probe experiments, including the edge-state distribution in the proximity of contacts.~\cite{Weitz00:247,Ahlswede02:165,Dahlem10:121305,Gerhardts13:073034} We would also like to bring attention to the fact that the screening theory is an approach to include electron-electron interactions at a fundamental level using a mean-field approximation, which is completely neglected in the approaches mentioned above. Recent works~\cite{Champel08:125302,Kotimaki12:053024,Oswald17:57009,Oswald17:125128,Armagnat20:02LT01,Werner20:235305,Kloss21:023025} utilize finite element and density functional theory methods that include the effect of the samples' finite sizes and the influence of contacts. These approaches are more advanced than the numerical methods used in screening theory. Their results well agree with the experiments and the approach presented here.

Our purpose here is to discuss the role of topology in real space, considering the edge and the bulk pictures of the IQHE. We discuss two geometries:
A Hall bar with identical square gates in the interior region and a Corbino disc as an annular 2DES surrounding a metallic contact (i.e., an electron reservoir (electrode)) and a second metallic contact surrounds the 2DES. We numerically obtained the electron density (filling factor) distribution and investigated the incompressible strip (IS) characteristics for both geometries with and without the disorder. In the case of a Hall bar with two inner contacts, we imposed an external non-equilibrium current and computed its spatial distribution with the Hall potential profile using the local version of Ohm's law.

Furthermore, we explicitly show that the compressibility of the bulk of the sample in real space plays no role in the observed quantization of the Hall potential. Instead, the quantization solely depends on the Hall probe contacts being decoupled from each other for the Hall bar, whereas for the Corbino geometry, the quantization is just a reflection of quantum capacitance between the source and drain contacts.

\section{Theory}

The topological treatment mentioned in the Introduction section requires a closed (periodic) and infinite (translationally invariant) sample geometry, namely a sample without edges. This geometry can only be realized in momentum space by considering a torus, a topological object with Genus number one.~\cite{Kane10:3045} However, to realize such a sample, in real space, one requires a line of magnetic monopoles aligned on a line centring the rotational axis of the geometry on providing a perpendicular magnetic field penetrating the sample's surface at every point. Unfortunately, no experimentalist has yet conducted charge transport measurements at a topological two-dimensional charge system with such sample configuration. In reality, on the one hand, at an ordinary Hall bar, one needs to drive a current (or apply a potential difference) between the source and drain contacts together with side probe contacts, which violates the assumed periodic boundary conditions together with the translational invariance. Hence, even within the linear response regime, the system is no longer in an equilibrium state for which one can claim periodicity and translational invariance. Moreover, adiabaticity is no longer valid with the electron transport velocities predicted~\cite{deniz07:075334} and measured.~\cite{McClure10:206806}

On the one hand, the samples have physical edges in real life regardless of whether they have current and probe contacts, i.e. if one tries to do measurements or not. The existence of physical edges was brought to attention by Halperin in his pioneering work demonstrating the importance of edge states.~\cite{Halperin82:2185} It is still under debate to understand that the current distribution in the QHE regime has a long history. Many transport experiments in connection with the quantum Hall effect is explained within the edge-state picture is developed by Büttiker,~\cite{Buettiker88:9375} elaborating on Landauer's quantum charge transport theory.~\cite{Landauer81:91}

On the other hand, an annular disc known as the Corbino geometry preserves the periodic boundary conditions. Laughlin's theory explains the quantized Hall conductance through a Corbino-disk geometry as discrete transfer of such quasiparticles between the inner and outer edges, one for each magnetic flux quantum threading the bore of the disk. Although the Corbino disks were suggested as a Gedankenexperiment initially, Dolgopolov and his co-workers realized such samples and measured Hall conductivity considering both Silicon MOSFETs and AlGaAs/GaAs materials,~\cite{Dolgopolov91:255,Dolgopolov92:12560} investigating the quantization in the absence of edge-states. We know that when the Hall conductivity is an integer multiple of $(e^2/h)$, the Fermi level is between two Landau levels, and the system is incompressible. A Corbino disc is the best electrode structure except for the non-uniform distribution of radial electric fields in the conducting ring when diagonal conductivity $\sigma_{xx}$ is uniform.~\cite{Yokoi98:249} The advantage of using Corbino disks is that the conductivity $\sigma_{xx}$ is directly measured, rather than obtained from inversion of the resistivity tensor.~\cite{deLang04:Phd} This is a crucial point to uncover transport which comprises both resistance and capacitance, i.e. the impedance.

Again, it is imperative to notice that these Corbino samples do not correspond to the original gedankenexperiment suggested by Laughlin, while in his theory, the magnetic flux is only changed at the bore of the cylinder, and the field is homogeneous elsewhere. On the other hand, Halperin's claimed topological equivalence between the cylindrical to the Corbino geometry only applies if the flux at the bore of the Corbino geometry is manipulated; however, the magnetic field seen by the electrons should remain constant and homogeneous. So, we hesitate that the experiments performed by Dolgopolov and his co-worker's experiments correspond to the same topology, although the geometry looks similar.

\subsection{First order non-interacting approach}
In a first-order approach, the two-dimensional electron system 2DES
consists of non-interacting electrons with no disorder. The
corresponding single-particle Hamiltonian is
given by
\begin{equation}\label{hamiltonian_landau}
H=\frac{1}{2m}(\mathbf{p}-e\mathbf{A})^{2},
\end{equation}
where $\mathbf{A}$ is vector potential generating the magnetic field $|\mathbf{B}|=B$ applied perpendicularly to the 2DES.

\subsubsection{The translationally invariant Hall bar}
The Hamiltonian can be solved in the Landau
gauge $\mathbf{A} = (0,Bx,0)$ 
\begin{equation}
\psi_{nk}(x,y)=\frac{1}{\sqrt{L}}e^{iky}\phi_{n}(x-X),
\end{equation}
where $X=-k^2\ell$ denotes the centre of the parabolic (magnetic) potential, with $k$ the quasi-continuous
momentum wave vector in $y$ direction, and the normalized eigenfunctions
\begin{equation}
\phi_{n,X}(x)=\frac{1}{\sqrt{\pi
ln!}}e^{-\frac{(x-X)^2}{2\ell^{2}}}H_{n}\left(\frac{x-X}{\ell}\right),
\end{equation}
depend on the Hermite polynomials $H_{n}$, $n\in \mathbb{Z}^{0+}$. Here, $\ell$ provides a quantum mechanical length scale $(=\sqrt{\hbar/eB}$) and is called the magnetic length. Although we used the Landau gauge, the symmetric gauge is more convenient for solving a disc geometry. Due to gauge invariance the
measurable physical quantities, i.e. energy eigenvalues $E_n$ and the electron density $n_{\rm el}(x,y)$, are the same apart from the complex pre-factor defining the phase of the wavefunction. It is important to recall that a translational invariance is assumed in the Landau gauge, whereas a rotational periodicity is assumed in the symmetric gauge. The first gauge yields a $y$ dependent quasi-momentum, defining a centre-coordinate. This centre coordinate (which has the dimensions of coordinate space) is a crucial parameter to normalize wavefunction and implicitly imposes that the particle number is conserved with and without external magnetic field within a specific area, namely, $\pi \ell^2$. By assuming above given boundary conditions and the Landau gauge, the corresponding energies of Landau levels are given as
\begin{equation}\label{landau_levels}
E_{n}=(n+1/2)\hbar\omega_{c},
\end{equation}
where
\begin{equation}
\omega_{c}=eB/m,
\end{equation}
is the cyclotron frequency of the orbit that charged particles with
mass $m$ and charge $e$ move in. The system has a highly degenerated density of states (DOS) since each Landau level can accommodate up to $eB/h=1/2\pi \ell^2$ electrons per unit area. The electron density divided by this quantity gives the number of Landau levels occupied by electrons and is called the filling factor $\nu$. In other words, the filling factor is the ratio of the number density of electrons to the number density of flux quanta:

\begin{equation}
\nu = \frac{n_{e}}{n_{B}}=2\pi\ell^2n_{e},
\end{equation}
where the pre-factor 2 reflects the fact that each Landau orbit can accommodate two spin species and the system is homogeneous, i.e. $n_{e}=n_{e}(x,y);\quad \forall x,y \in(-\infty,\infty)$, where $n_{e}(x,y)$ is the local electron density, hence the local filling factor $\nu(x,y)=2\pi\ell^2n_{e}(x,y)$, in real space.

Moreover, the filling of Landau levels can be
combined with the drift velocity. The Hall current $I_{\rm H}$ is then
expected to be
\begin{equation}\label{hall_current}
|I| = n(e^2/h)|E|,
\end{equation}
with $n$ being the number of filled Landau levels and $I_{\rm H}$ is in
the direction perpendicular to driving electric field $\mathbf{E}$. It follows that the Hall
conductivity is an integer multiple of $(e^2/h)$:
\begin{equation}
\sigma_H = n(e^2/h),
\end{equation}
when the Fermi level is between two Landau levels, i.e. if the system is incompressible.
These relatively simple, however fundamental, formulas connect the Hall conductivity linearly with the
quantum number $n$. However, explaining the high reproducibility of the precise plateaus and their finite widths are far too complicated than this simple argumentation. 

By imposing relatively reasonable realistic boundary conditions in $x-$ direction, say infinite potential walls~\cite{Halperin82:2185} or smoothly varying confinement potential,~\cite{Buettiker86:1761,Chklovskii92:4026,Oh97:13519,Guven03:115327,Siddiki04} one ends up with so-called edge states carrying the equilibrium current. On the one hand, in some of these approaches, actual metallic contacts where the potential difference or current is imposed are assumed to be responsible for all non-equilibrium processes. Namely, the dissipation and equilibration occur at the contacts, whereas edge-states are free from scattering (ballistic), and transport of non-equilibrium current can be described by Landauer formalism.~\cite{Landauer81:91} It is essential to emphasize that the edge-states are nothing but the superpositions of the Landau wavefunctions at the Fermi energy with the same centre coordinate, $X$.~\cite{Gerhardts1} On the other hand, a dissipative Hall bar is considered by R. R. Gerhardts and his co-workers,~\cite{Guven03:115327,Siddiki04} which is in perfect agreement with the local probe experiments conducted by J. Weis and his co-workers.~\cite{Weitz00:247,Ahlswede02:165,Dahlem10:121305,Gerhardts13:073034} Moreover, the dissipative Hall bar, including hot spot phenomena, is explicitly calculated by Siddiki and co-workers~\cite{Kilicoglu10:165308,GUVENILIR19:283} utilizing numerical methods, showing that it is not necessary to assume ideal contacts (where the contacts absorb all the current without dissipation). Their findings are also in perfect alignment with the above mentioned local probe experiments, considering the screening properties and also the capacitive behaviour of the Hall bar.~\cite{Suddards12:083015}

\subsubsection{The annular disc geometry, the Corbino device}
The quantized Hall conductance at a Corbino-disk geometry can be understood as a discrete transfer of a two-dimensional electron between the inner and outer edges, one for each
magnetic flux quantum threading the bore of the disk.\cite{Laughlin81:5652} In addition, note that the above argumentation relies on the fact that the Hall conductivity is quantized as an integer multiple of $(e^2/h)$. The Fermi level is between two Landau levels and, therefore, the system is incompressible, i.e. no available electronic states at the Fermi level. Moreover, disorder emanating from remote donors provides localized states, resulting in quantized Hall plateaus of finite widths at certain $B$ field intervals and keeping the incompressibility of the sample. Namely, without the disorder, the plateaus would shrink to a single $B$ value where the Landau level filling $\nu$ is an integer, which is far different from the observations both in Hall and Corbino geometries. 

The first and significant difference between the Hall bar and the Corbino geometry is the orientation of the edge-states to equilibrium current direction and edges. At the Hall bar geometry, the edge-states are perpendicular to the contacts. However, at a Corbino device, they are parallel to the contacts. Hence, one cannot simply apply the same equilibrium edge-state picture to describe the charge transport through the inner and the outer contacts. Here, a wavefunction dependent transport model becomes indispensable, which Halperin also introduced.~\cite{Halperin82:2185} In his seminal work, Halperin extends the edge-state picture by including disorder and discussing the existence of extended states (edge-states) in two-dimension and concludes that, without the disorder, the current can be described as
\begin{equation}
I_{j,n}\simeq\frac{e^2B_0}{m}\int_{0}^{\infty}dr |\psi_{j,n}|^2(r_j-r),
\end{equation}
where the symmetric gauge is used, $\mathbf{A}=\frac{1}{2}B_0 r+\Phi/2\pi r$, $\mathbf{B_0}$ is perpendicular to the disc surface, and $\Phi$ is the amount of magnetic flux. The radius of $r_j$ is determined by,
\begin{equation}
B_0\pi r_j^2=j\Phi_0-\Phi,
\end{equation}
$\Phi_0$ being the quantum flux, $hc/e$. The integer quantum number $j$ (note that in the original paper $m$ is used as the quantum number) translates itself to quasi-momentum in $y-$ direction $k$, in the Landau gauge. Whereas, for the disordered system, the expression for the wavefunction is modified using localization and extended state arguments, which results in a finite plateau width.

It is crucial to recall by his own words that "\emph{We assume in addition that there is a magnetic flux $\Phi$, confined to the interior of a solenoid magnet threading the hole in the annulus, and we shall be able to vary the flux $\Phi$ without changing the magnetic field in the region where the electrons are confined. (This is a slight modification of the cylinder geometry considered by Laughlin. ) We shall assume that no electric field is present so that the electrostatic potential seen by
the electrons is constant in the interior of the film, and we assume that the dimensions of the annulus are very large compared to the cyclotron radius $r_c$, for electrons in the magnetic field.}". The above assumptions are far different from what is conducted in experiments by Dolgapolov and others.~\cite{Dolgopolov91:255,Dolgopolov92:12560,deLang04:Phd}  

In Corbino geometry experiments, the conductance is measured as a function of the homogeneous $B$ field applied or by changing the electron density with the help of a constant potential (electric field) applied to metallic gates. Namely, the field inside the bore is also changing, and there exists an electric field seen by the electrons. Hence, either way, the conductance is measured as a function of the filling factors. Recall that these experiments are essentially two-point measurements, different from Hall bar experiments. 

This inconsistency was already realized by Halperin reporting that "\emph{In a real experiment, the measured Hall potential $eV$ is the sum of an electrostatic potential $eV_0$ and
the difference in Fermi levels $E_F^{(2)}-E_{F}^{(1)}$. The edge
current is then only a \emph{fraction} of the total Hall
current, given by $(E_F^{(2)}-E_{F}^{(1)})/eV\approx \alpha n r_c \hbar \omega_c /e^2$
where $C$ is the capacitance per unit length of the
edge states, and $\alpha$ is a number of order unity.}" An explicit expression capacitance is not given in the paper, however, is suggested by himself to be a quantum capacitance,~\cite{Halperin:private} which is directly connected to the compressibility of the system.~\cite{Kilicoglu16:035702}

In the following subsection, we introduce the fundamental formalism to calculate self-consistently both the electron and current density distributions assuming periodic boundary conditions in both directions, within the Thomas-Fermi approximation (TFA) that assumes that the potential varies smoothly on quantum mechanical length scales.

\subsection{The self-consistent approach}
The inclusion of interactions to Hamiltonian given in Eqn.~\ref{hamiltonian_landau} is far too complicated to be solved by exact analytical or numerical methods for actual sample geometries.~\cite{Girvin90:book,Ezawa00:book,Yoshioka02:book} In the early days of the IQHE theory, interactions were utterly ignored since the interparticle distance (i.e. the Fermi wavelength, $\lambda_F$) without a magnetic field is at the order of 30-40 nm and the effective screening length (i.e. effective Bohr radius $a^*_B$ is at the order of 10 nm, for AlGaAs/GaAs samples with average electron densities $n_{el}\approx 1-3\times 10^{-11}$ cm$^{-2}$. Hence, it was reasonable to consider only a single particle Hamiltonian. However, the neglection of interactions was an underestimation of the long-range behaviour of the direct Coulomb interaction. This long-range effect can be considered at the lowest order correction to the single-particle Hamiltonian via momentum dependent dielectric function $\epsilon(q)$ (neglecting time variation of the external electric field). It can be calculated utilizing the Thomas-Fermi type mean-field approximation.~\cite{Ashcroft} In this manner, screening theory finds its firm ground via the relation

\begin{equation}\label{dielec}
V^q=\frac{V^q_{\rm ext}}{\epsilon(q)},
\end{equation}
where $\epsilon(q)$ directly depends on the density of states (DOS), which strongly depends on the dimension and the quantization due to the magnetic field.~\cite{Siddiki03:125315} This fact reflects itself in non-linear screening and gives rise to the formation of compressible and incompressible (local) regions.~\cite{Chang90:871,Chklovskii92:4026,Chklovskii93:12605, Siddiki03:125315} Also note that, quantization effect due to $B$ field is already included implicitly via DOS.

\subsubsection{Equilibrium calculations}
Instead of seeking an exact many-body solution, we employ the lowest order, however well-appreciated, density functional approximation. Namely, the Thomas-Fermi-Poisson approximation (TFPA) to obtain the electron density and potential distributions self-consistently which solves the Poisson equation numerically for given boundary conditions.~\cite{Lier94:7757,Guven03:115327,Sefa08:prb} The main advantage of this approximation is the fact
that the Dirac-delta (function) replaces the Landau wave functions, i.e. $|\phi_{n,X}|^2\approx\delta(x-X)$, and the center
coordinate dependent eigenvalues can be given in the lowest order
perturbation as
\begin{equation}
E_{n}(X,y)\approx E_{n} + V(X,y).
\end{equation}
The TFPA allows one to surpass the numerical complications which may arise due to the solution of the Schrödinger equation if the total (screened potential) varies slowly on quantum mechanical length scales. However, the quantum mechanical solutions within the Hartree approximation are reported in the literature considering a translational invariant geometry in the current direction, which yield similar results to our approach presented here.~\cite{Suzuki93:2986,siddiki2004}

We calculate the electron density and the total potential (energy) from the following self-consistent equations:
\begin{equation}\label{nelec}
n_{\rm el}(x,y)=\int
dE\frac{D(E,(x,y))}{e^{[E+V_{\rm tot}(x,y)-\mu_{\mathrm{elch}}^{\star}]/k_{B}T}+1},
\end{equation}
and
\begin{equation}\label{vpot}
V_{\rm tot}(x,y)=V_{\rm ext}(x,y)+V_{\rm int}(x,y),
\end{equation}
where $V_{\rm ext}(x,y)$ is the applied external and $V_{\rm int}(x,y)$ is
the Hartree interaction potential (energy) between the electrons in the system defined as
\begin{equation}
V_{\rm int}(x,y)=\frac{2e^2}{\overline{\kappa}}\int
K(x,y,x^{\prime},y^{\prime})
n_{\rm el}(x^{\prime},y^{\prime})dx^{\prime}dy^{\prime}.
\end{equation}
Here $\mu_{\mathrm{elch}}^{\star}$ is the electrochemical potential which is constant
in the absence of an external current $I$, $\overline{\kappa}\quad(\sim12.4$ for GaAs/AlGaAs heterostructures) is an average
dielectric constant and $K(x,y,x^{\prime},y^{\prime})$ is the solution of the Poisson equation considering periodic boundary conditions.~\cite{Morse-Feshbach53:1240}
Note that the electrochemical potential in equilibrium is defined as
\begin{equation}
\mu_{\mathrm{elch}}^{\star}=\mu_{\rm ch}-|e|\phi(x,y)
\end{equation}
where $\mu_{\mathrm{\rm ch}}$ is the chemical potential and $\phi(x,y)$ is the electrostatic potential.
In addition, the chemical potential is determined by the statistical description assuming
a grand canonical ensemble which is in contact with a reservoir.

\subsubsection{Imposing a non-equilibrium current}
We obtain $n_{\rm el}(x,y)$ iteratively in
thermal equilibrium, keeping the donor distribution fixed
($n_{0}$) and average electron density constant, i.e. $\mu_{\rm ch}$ is constant and position independent, starting from
$T=0$ and $B=0$
solutions. Similar calculations are already reported in the literature.~\cite{Guven03:115327,siddiki2004,SiddikiMarquardt,Sefa08:prb,Kilicoglu10:165308}

In the presence of a fixed external current driven in the
longitudinal ($y$-) direction the scheme described above becomes rather
complicated. Since, the external current modifies $n_{\rm el}(x,y)$ and $V_{\rm tot}(x,y)$, therefore should be
computed once more self-consistently. By assuming a local thermal
equilibrium~\cite{Guven03:115327}, the driving electric field is given by the gradient
of the position dependent electrochemical potential,
\begin{equation}
\mathbf{E}(x,y)=\nabla\mu_{\mathrm{elch}}^{\star}(x,y)/e=\hat{\rho}(x,y)\mathbf{j}(x,y),
\end{equation} and for a given
local resistivity tensor
($\hat{\rho}(x,y)=[\sigma(n_{\rm el}(x,y))]^{-1}$), one can obtain the
position-dependent electrochemical potential. This $\mu_{\mathrm{elch}}^{\star} (x,y)$
will be used to obtain the new $n_{\rm el}(x,y)$ and $V_{\rm tot}(x,y)$ in
the next iteration step. The detailed description of the
method can be found in Ref.~\cite{Guven03:115327}.\\
\indent We assume that the charge density changes slowly on the correlation length
of the remote impurities, which allows us to obtain the conductivities from the self-consistent
Born approximation (SCBA).~\cite{Ando82:437} Local conductivities are assumed to be directly related to the local electron density.~\cite{Guven03:115327}
The explicit forms of the conductivity tensor elements for a homogeneous system are discussed in detail in Refs.~[\onlinecite{Ando82:437,Guven03:115327,siddiki2004}].
Having the local electron density and the local magneto-transport coefficients at hand, we perform calculations to obtain the current distribution utilizing the above
described microscopic model, assisted by the local Ohm's law at a fixed external current
$I$. Further details of the calculation scheme are reviewed in Ref.~[\onlinecite{Gerhardts08:378}]. Note that, since our results are independent of the particular nature of the single-particle gap, the spin-splitting of
the Landau levels will be neglected, and the spin degeneracy will be taken into account by the degeneracy factor $g_{s}=2$ throughout this study.

\section{Results and Discussions}
As announced in the Introduction, our purpose is to discuss the role of interactions on the topological insulators of the quantized Hall effect, namely the incompressible states. To do so, we will investigate two different geometries, namely, the transitionally invariant geometry: The Hall bar and a rotationally symmetric geometry: The Corbino disk, at particularly interesting magnetic field regimes. By performing self-consistent calculations at particular magnetic fields that characterize the system behaviour, we show that the topological insulators' bulk and edge representation merge in a strong disorder potential. Namely, the momentum and coordinate space duality are re-constructed with localization arguments. Therefore we focus on the spatial distribution of the formation of the incompressible regions in both geometries depending on magnetic field strength. To be explicit, we will investigate whether the incompressible region resides at the bulk or the edge and how they merge.

\subsubsection{The Hall bar with two inner contacts and current driving electric-field}

The self-consistent solution of coupled equations (\ref{nelec}) and (\ref{vpot}) within the Thomas-Fermi approximation provides precise results concerning the density and potential profiles of the 2DES under high magnetic fields. In addition, TFPA enables us to determine the locations of the incompressible strips at the edge or bulk of the Hall bar and around the contacts. In our first step of the investigation, we assume periodic boundary conditions and ideal source and drain contacts to be consistent with previous approaches. However, it is essential to note that we impose periodicity to the total potential, i.e., the interaction (screened) and imposed voltage difference at the contacts, enabling us to drive a non-equilibrium current.

\begin{figure}[t!]{\centering
\includegraphics[width=1\linewidth]{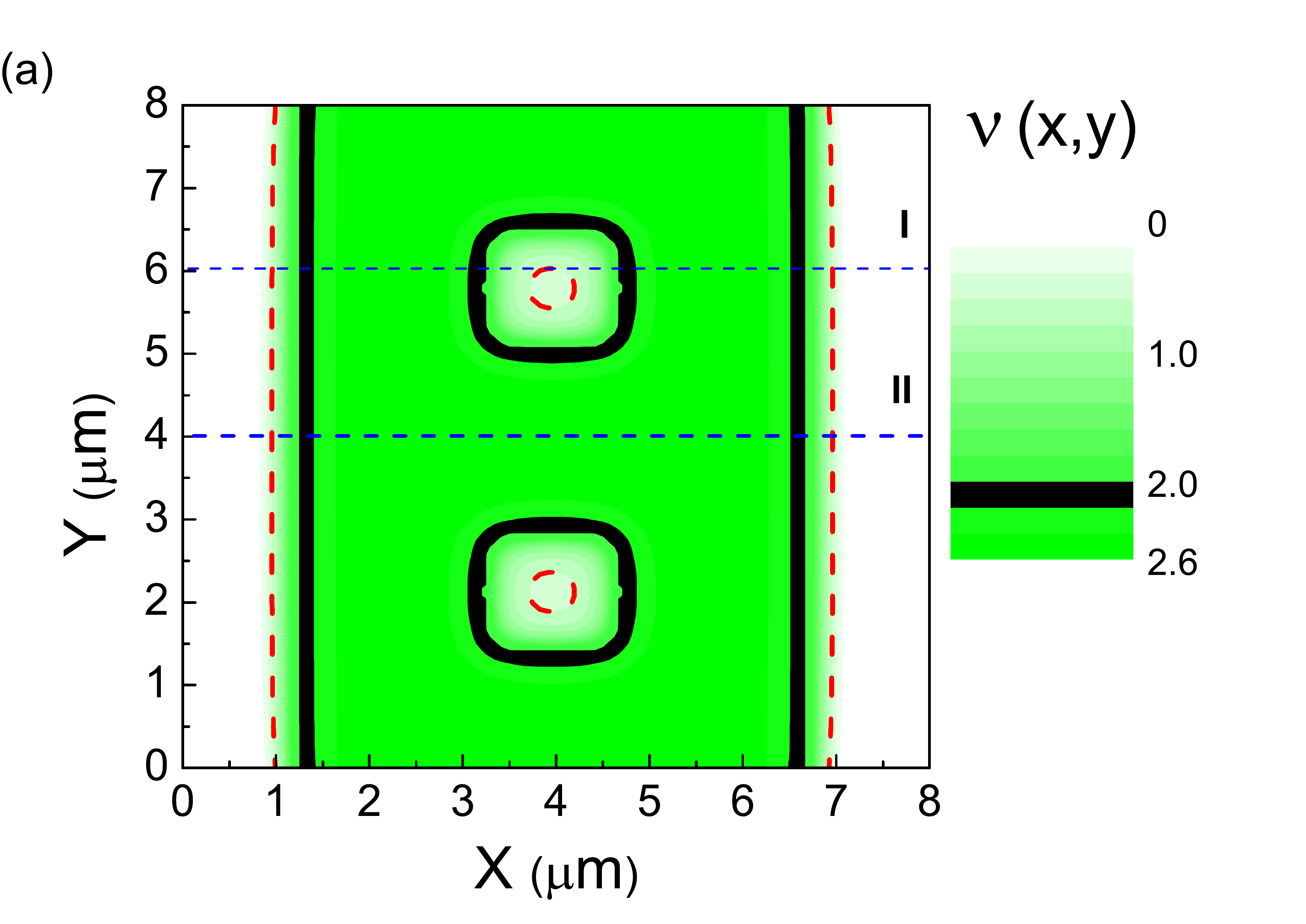}
\includegraphics[width=1\linewidth]{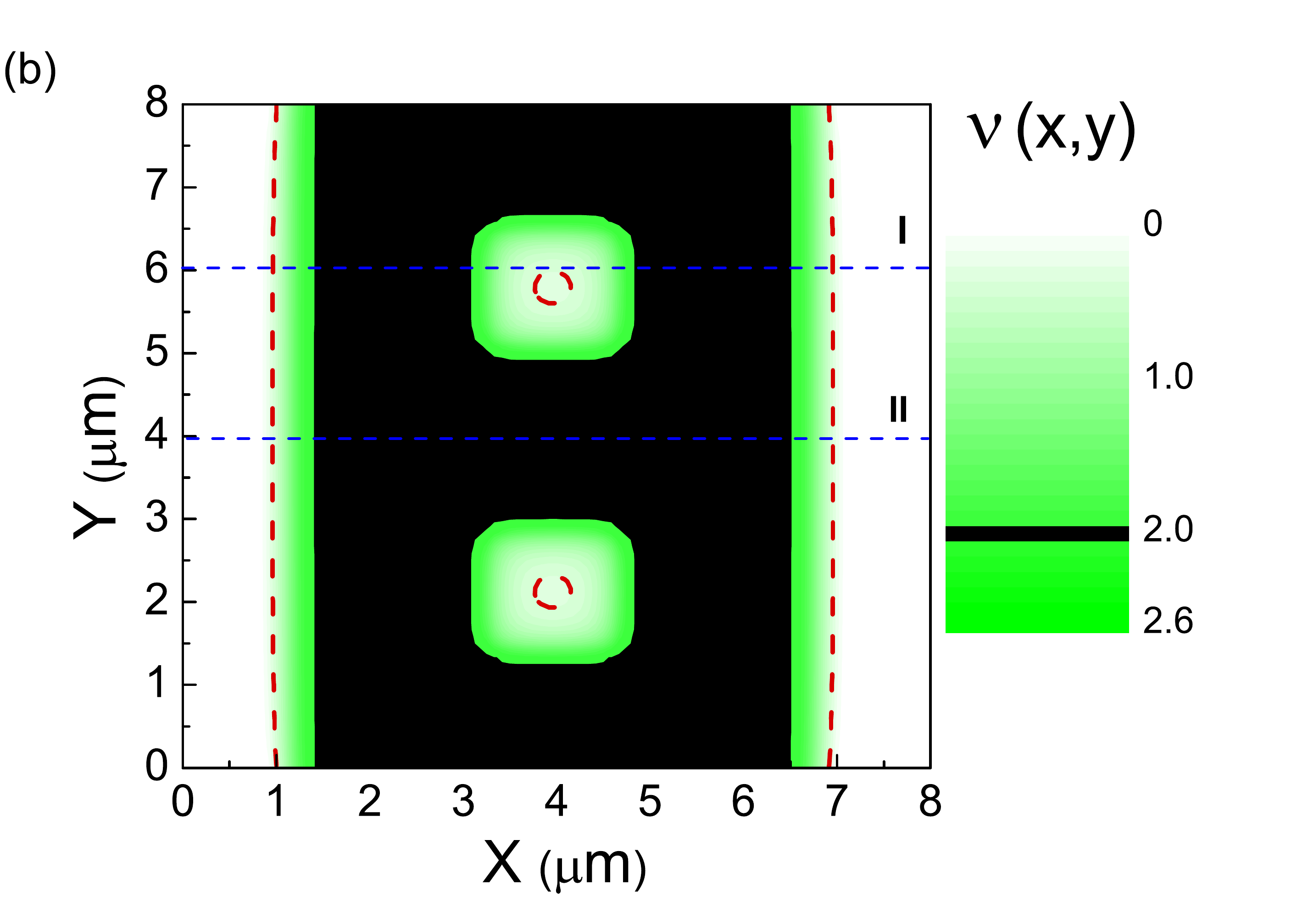}
\vspace{-0.5cm}\caption{\label{fig:hall_bar} (Color online) (a) The local filling factor distribution of a disorder free Hall bar at $B=7.5$ T and $T=6$ K. (b) Same quantity at a higher field $B=9.1$ T. Grey (green)-scale depicts the density gradient, whereas black regions correspond to $\nu=2$ incompressible regions. Source and drain contacts are at the ends of the sample in $y-$ direction and are scatterring free, i.e. without density gradient in their proximity. In contrast inner contacts induce a density gradient near them, surrounded by incompressible strips.}}
\end{figure}

The distribution of the local filling factors $\nu(x,y)$ at a fixed magnetic field of $B=7.5$ at a finite temperature of $T=6$ K is shown in Fig.\ref{fig:hall_bar}a. We observe that the interior of the Hall bar is mostly compressible, depicted by light grey-scale (green) regions. The electron depleted regions are white and countered with (red) broken lines. The incompressible (emphasized by black colour) strips are located along the physical edges of the sample and in the surrounding proximity of inner contacts. Whereas, at a higher magnetic field of $B=9.1$ T and considering at $T=4$ K, the Hall bar becomes entirely incompressible in Fig.\ref{fig:hall_bar}b. Except for the proximity of inner contacts, as expected. We observe that changing the $B$ field alternates between the edge picture, i.e. Landauer-Büttiker, and the bulk picture, i.e. Laughlin's theory, without changing the system's quantization properties.

Meanwhile, having two inner contacts destroys the topology of genus 1 to genus 3 in real space. One can argue that having such inner contacts correspond essentially to the Corbino geometry, which is not correct since, as we will discuss in the following subsection, we are not applying a potential difference between these two inner contacts or between the inner contacts and the outer, source/drain contacts. These two inner contacts can be supposed to be taken as two isolated islands where no electrons exist. As shown experimentally, the existence of these two contacts does not affect the QHE, even in the case of fractional states.~\cite{Kendirlik17:14082} For an alerted reader, we refer to our previous work discussing in detail the electron density distribution within the Hall bar comprising inner contacts and disorder. However, the system's physical parameters are slightly different, and no external current is considered.~\cite{Yildiz14:014704}

\begin{figure}[t!]
\includegraphics[width=1\linewidth]{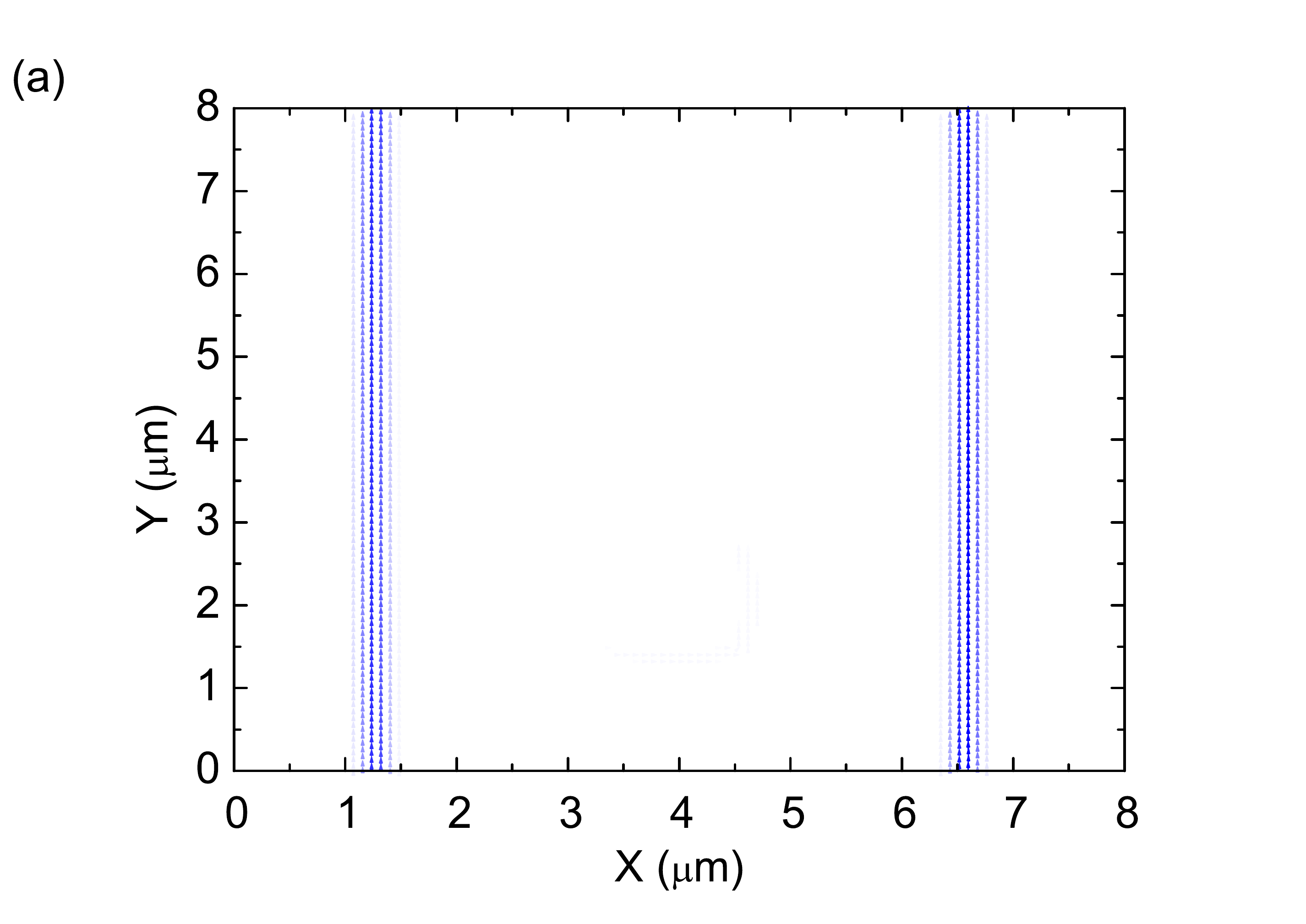}
\includegraphics[width=1\linewidth]{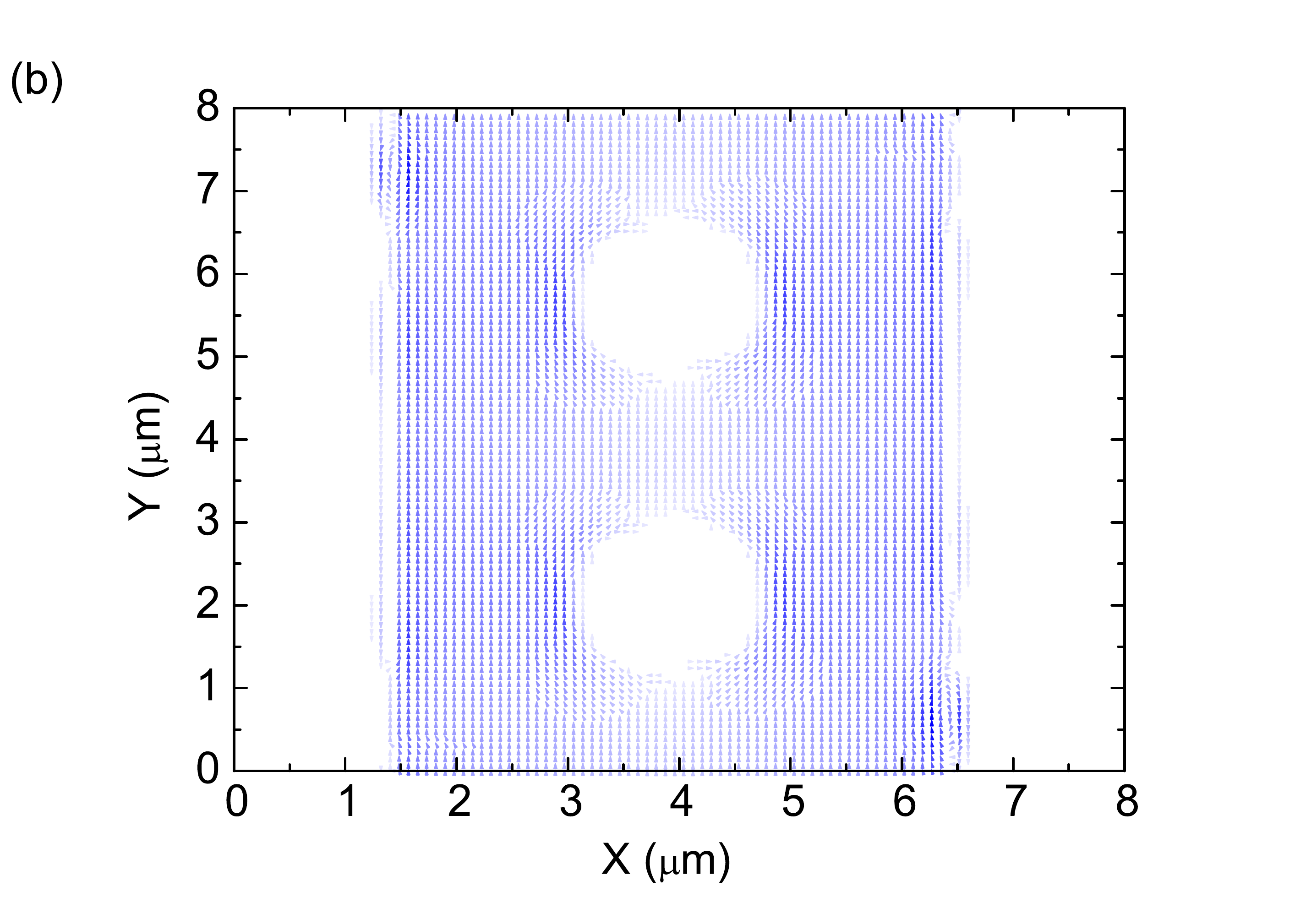}
\caption{\label{current_edge} )Color online) (a)
The current distribution in the dissipative Hall bar at $B=7.5$ T and T=6 K and (b) at $B=9.1$ T. The arrows indicate the current direction, whereas the grey (blue)-scale depicts the current intensity. The total current density applied in $y-$ direction is $2.24\times 10^{-3}$ $\rm A/m$, corresponding to $\sim 18$ nA including the electron depleted regions at the physical boundaries. Left upper and right lower regions with increased current density indicates the formation of hot-spots even considering ideal contacts.}
\end{figure}

To obtain the current and position-dependent electrochemical potential distributions in the presence of additional inner contacts, a sufficiently weak electric field is applied in the longitudinal direction that drives a negligible current density of $2.24\times 10^{-3}$ $\rm A/m$.
In this case, the linearity between current density and the electric field is conserved to a good approximation; hence, the linear response approximation is still valid.~\cite{siddiki2004}
The current distribution obtained within the linear response regime is shown in Fig.~\ref{current_edge}. The one-to-one correspondence between the spatial distribution of the incompressible stripes and the current paths can be seen if one compares Fig.~\ref{fig:hall_bar} and Fig.~\ref{current_edge}, except that no non-equilibrium current is flowing from the incompressible stripes surrounding the inner contacts. In contrast, all the imposed non-equilibrium current is confined to the edge incompressible strips since the source and drain contacts are ideal. The non-equilibrium current does not flow from these strips since only the edge stripes are in direct contact with the reservoir (injection contacts), and since the incompressible stripes are dissipation-less, the current is not transferred to the bulk. Fig.~\ref{current_edge} manifests the scattering free property of the incompressible stripes, where the external non-equilibrium current is confined. In addition, the current distribution considering the edge configuration presents a slight asymmetry, such that the right edge carries more current indicated by the darker colour to the left edge.

This slight difference indicates that even considering a relatively small current at the order of a few $\mu$A can influence the density distribution~\cite{Guven03:115327} and an idealized linear response is not possible in the IQHE regime, as discussed previously.~\cite{Sefa08:prb,Siddiki09:17008} This argumentation is also evidenced by the experiments performed on high mobility narrow gate defined Hall bars~\cite{SiddikiNJP10:113011} and by scanning force microscopy measurements.~\cite{Ahlswede02:165} Recently, such an asymmetric current distribution has been investigated experimentally and theoretically in detail.~\cite{Gerhardts13:073034} The current is carried by the edge incompressible stripes at $B=7.5$ T, while by increasing the magnetic field up
to $9.1$ T, the current starts to flow from the whole bulk of the sample, Fig.~\ref{current_edge}b, avoiding the inner contacts where no electron exists. A careful analysis of the current distribution for the bulk case shows a relative local increase of the current (and also charge density) in the close vicinity of the inner contacts. This observation perfectly coincides with the findings of Gross and Gerhardts, where they have investigated the current distribution near the breakdown of the IQHE, considering two anti-dots at the bulk of the sample.~\cite{Gross98:60} The anti-dot geometry is analogous to the system we consider; however, the formation of incompressible stripes is not considered in the anti-dot case. In light of our above observations, we expect to obtain a slight increase in the resistance between the inner contacts once the incompressible strips start to form either at the higher or lower edge of the Hall plateau, hence scattering from and to the compressible to incompressible regions become possible therefore the linear increase in $B$ should be interrupted.

The current flow from the edge incompressible stripes points to a possible finite current flow between inner contacts since the bulk is compressible. This situation does not affirm Halperin's edge picture, where the bulk of an annular disc is entirely incompressible.~\cite{Halperin82:2185} Clearly, the incompressible strips surrounding the inner contacts would yield a capacitive barrier. However, at sufficiently large potential differences between the inner contacts, it is in principle possible to drive a finite current. On the other hand, the one-to-one correspondence with Halperin's picture can be obtained at the high-$B$ edge of the quantized Hall plateaus, or if an intense disorder is imposed, we will discuss in the following section considering a Corbino geometry. The current distribution of a disorder free situation, Fig.~\ref{current_edge}b, indicates that if the bulk becomes completely incompressible, it is not possible to inject current between the inner contacts due to reasonably high impedance,~\cite{Yildiz14:014704} which is already experimentally confirmed and reported recently.\cite{Kendirlik17:14082} However if the resistance is measured independently, one would probably observe a quantized behaviour regardless of which plateau one performs the measurements.
\begin{figure}[t!]
\includegraphics[width=1\linewidth]{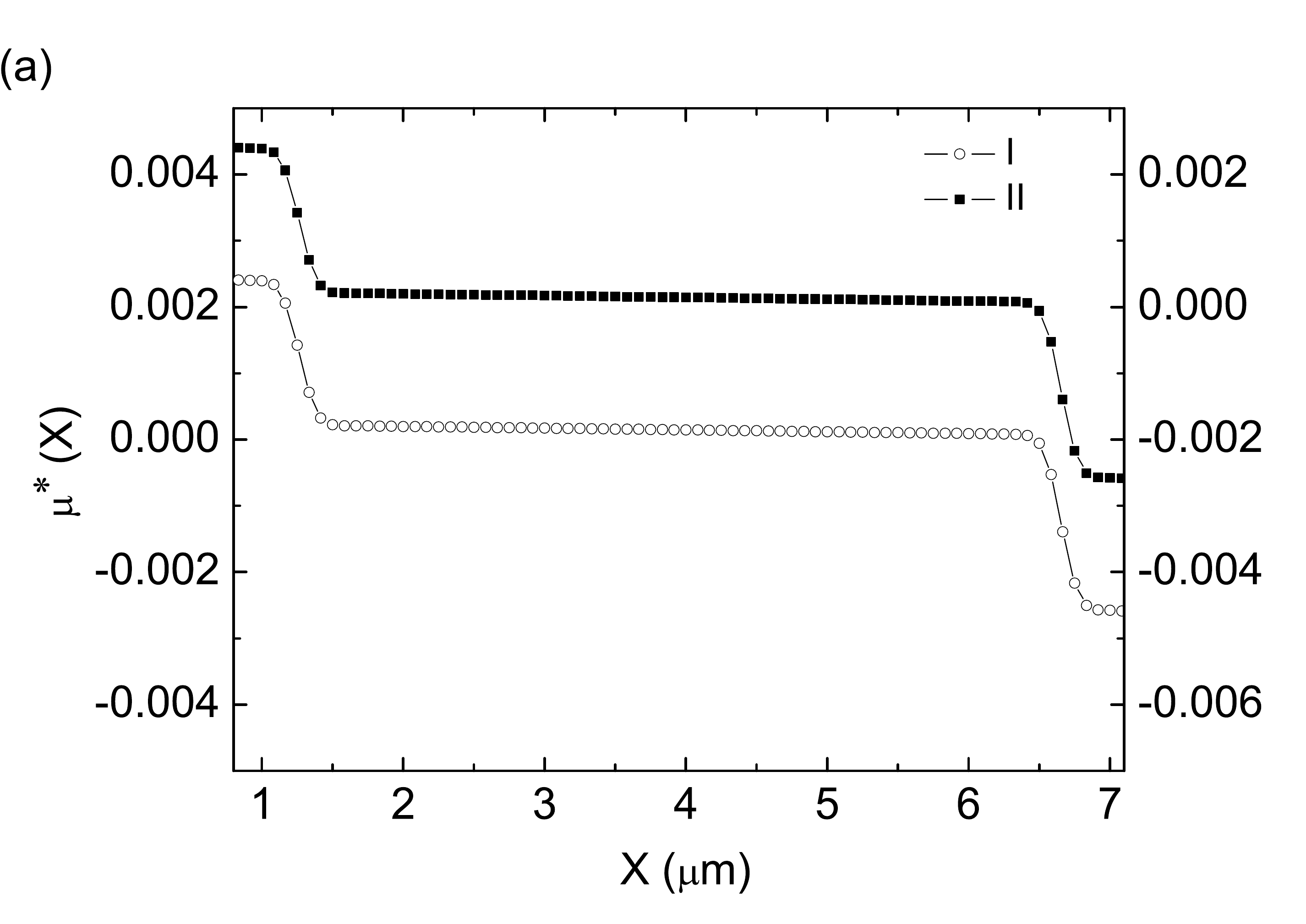}
\includegraphics[width=1\linewidth]{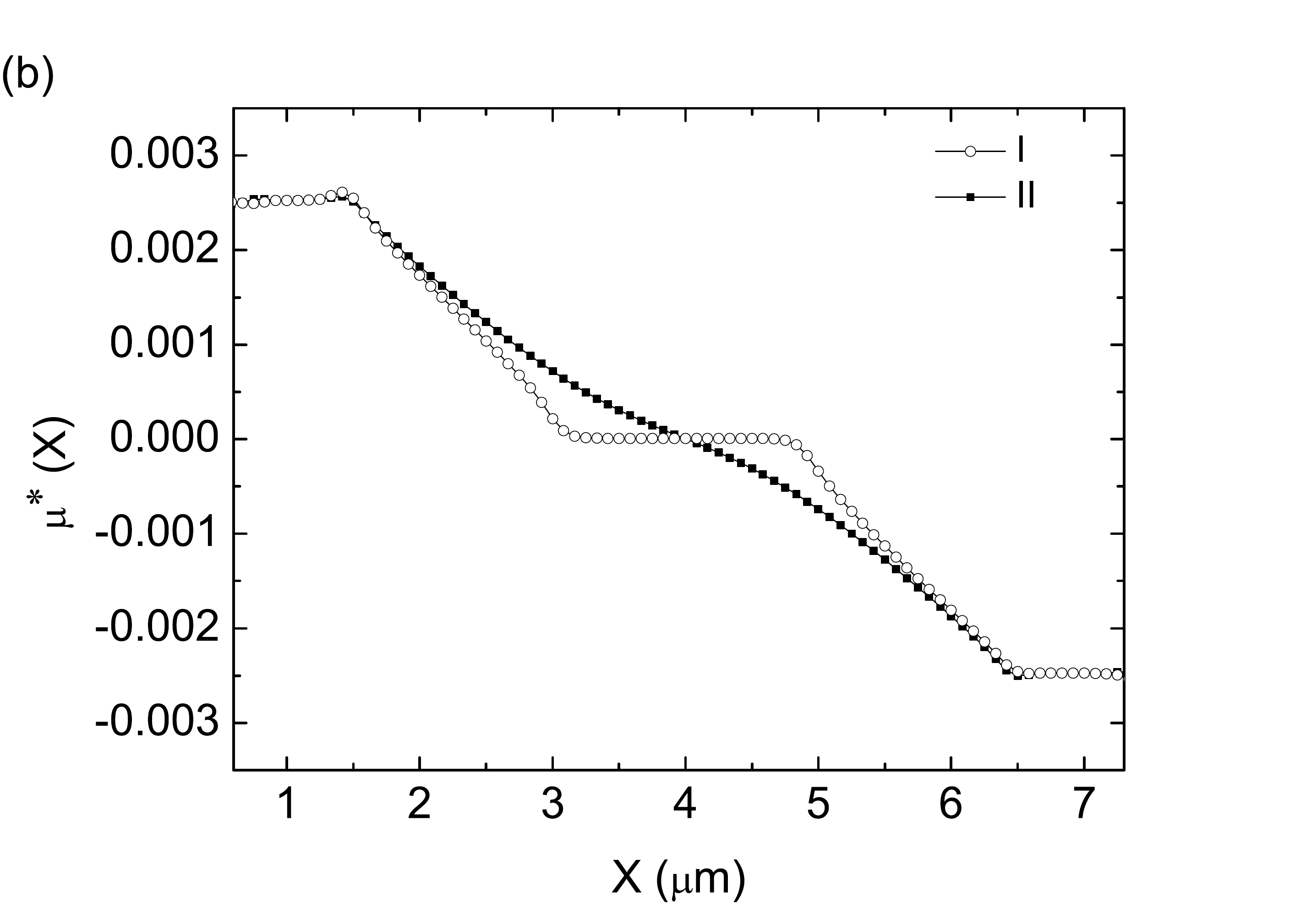}
\caption{\label{hallbar_edge_elchempot}The spatial distribution of the electrochemical potential regarding the Hall bar geometry assuming the parameters given in Fig.~\ref{current_edge} at (a) low and (b) high field limits (b). In Fig.~\ref{hallbar_edge_elchempot}(a) $y$ values are shown by two scales on both sides to avoid overlapping curves. Left corresponds to cross-section I and right to cross-section II. One can see that the imposed small current does not considerably affect the total potential. Hence, the system can be regarded as to be in the linear response regime.}
\end{figure}

We also plot the electrochemical potential distribution, $\mu_{\mathrm{elch}}^{*}(x,y)$, considering
both edge and bulk states at different cross-sections in $y-$ indicated by horizontal broken lines in Fig.~\ref{fig:hall_bar}, marked by I and II. The self-consistently obtained $\mu_{elch}^{\star}(x,y)$ is plotted in Fig.~\ref{hallbar_edge_elchempot} as function of lateral coordinate $x-$. It is observed that the potential varies exactly at the incompressible edge strips. Since these strips are scattering free and carry a dissipative non-equilibrium current uni-directionally. Note that dissipation is exponentially small at low temperatures and vanishes in the limit of $T\rightarrow0$. This result is in excellent agreement with the scanning probe experiments~\cite{Weitz00:247,Ahlswede02:165,Dahlem10:121305,Gerhardts13:073034} and theoretical predictions.~\cite{Guven03:115327,Siddiki04:195335} Most interestingly, the electrochemical potential distribution is unaffected by the inner contact(s), which can be seen by tracing the curve corresponding to the location I (cf. Fig~\ref{hallbar_edge_elchempot}a, upper curve). In other words, since the inner contacts are allowed to float, they are in equilibrium with the compressible region surrounding them. Also, notice that no current at the bulk can flow in the transverse direction (i.e. Hall direction) since there is no driving electrical field (due to good screening, the potential is flat) is in this direction. Namely, the electrochemical potential is constant at the entire bulk. The only potential drop occurs at the locations where incompressible strips reside. Hence, Hall potential (electric field) is confined to these strips. An increase of the external $B$ field to 9.1 T results in the formation of an incompressible bulk region as shown before, and the corresponding electrochemical potential distribution is plotted in Fig.~\ref{hallbar_edge_elchempot}b. It is interesting to observe that, once the bulk becomes incompressible, the potential distribution is strongly affected by the inner contacts, such that the electrochemical potential varies across the sample where there is no contact. However, it remains constant within the inner contact if the trace of the $\mu_{\mathrm{elch}}^{\star}$ is calculated at location II as shown in Fig~\ref{hallbar_edge_elchempot}b, lower curve. Such a difference also coincides with early experiments probing the Hall potential distribution.~\cite{Zheng85:5506} However, a quantitative comparison is not realistic since a low-mobility large sample is used in the experiments, whereas our calculation considers a narrow disorder free system. In any case, this potential profile imposes that if one drives current between these two inner contacts, there would be a finite potential difference (in fact, due to high quantum capacitance); hence the impedance measured between these contacts would be considerably large, as reported.~\cite{Yildiz14:014704,Kendirlik17:14082}

To summarize, we calculated a finite size Hall bar's density, potential, and current distributions by imposing periodic boundary conditions on the total potential and assuming ideal source and drain contacts. In addition, we located two inner contacts inside the sample, which altered the system's topology in real space and observed that the current distribution and resistance quantization remains unaffected, in contrast to bulk pictures of the IQHE theories. However, our results do not exclude the bulk picture entirely but show that the topological argumentations should be re-examined while the coordinate-space and momentum-space duality breaks down if one considers finite-size effects and current/voltage contacts as is actual experiments does have.

In the next section, we will investigate a Corbino geometry and compare the spatial distributions of the incompressible regions to the Hall bar geometry. We will show that, as expected, in contrast to Hall bar, at Corbino geometry, the edge-states are no longer perpendicular to the contacts. However, the quantization is obtained by the exact mechanism that decouples the probe contacts due to incompressible states. In addition, we will also investigate the effect of disorder at the bulk of the sample, mainly, and show that similar to Hall bar geometry, edge and bulk pictures merge in the real space.

\subsubsection{The Corbino disc without driving electric field}

As mentioned briefly in the introduction, Dolgopolov and his coworkers reported one of the earliest realizations of Corbino geometry conductance measurements.~\cite{Dolgopolov91:255,Dolgopolov92:12560} One of the most exciting results reported was that no quantization was observed at low mobility SiGe MOSFET, in contrast to relatively well-developed conductance plateaus observed at AlGaAs/GaAs samples, and they concluded that this observation could not be explained with experimental errors. Another critical investigation conducted at the Corbino devices that concerns the present manuscript was performed at von Klitzing's group to understand the tunnelling between the edge channels.~\cite{Liu98:4028} From the theoretical aspect, a similar approach was used to determine the local conductivities from local electron densities. It was concluded that a self-consistent model and a quantum mechanical tunnelling formalism are required to understand the current-plateau observed.

A few years after the above investigation, Corbino disks were fabricated in the same group with few microns in size and investigated by scanning force microscopy techniques.~\cite{Ahlswede02:165} Experimentally, it was shown that within a conductance plateau, the electronic system comprises an incompressible region, which can be at the bulk or the edges as stripes. The incompressible bulk state is in perfect agreement with Laughlin's argument, besides an electron-poor region in front of the contacts,~\cite{Kilicoglu10:165308,Dahlem10:121305} however, the observed stripes encircling the edges of the Corbino sample became quite confusing. Since the bulk remains compressible in the presence of edge incompressible stripes, whence electrons can be transferred from outer contact to inner contact without phase coherence arguments (i.e. a finite scattering) at the bulk. However, the Hall conductance remains quantized even in this case, as evidenced by the experiments.~\cite{Ahlswede02:165,Dahlem10:121305} Here, we will also focus on this phenomenon and provide detailed numerical results considering a Corbino disc with and without the disorder.

\begin{figure}[t!]{\centering
\includegraphics[width=1\linewidth]{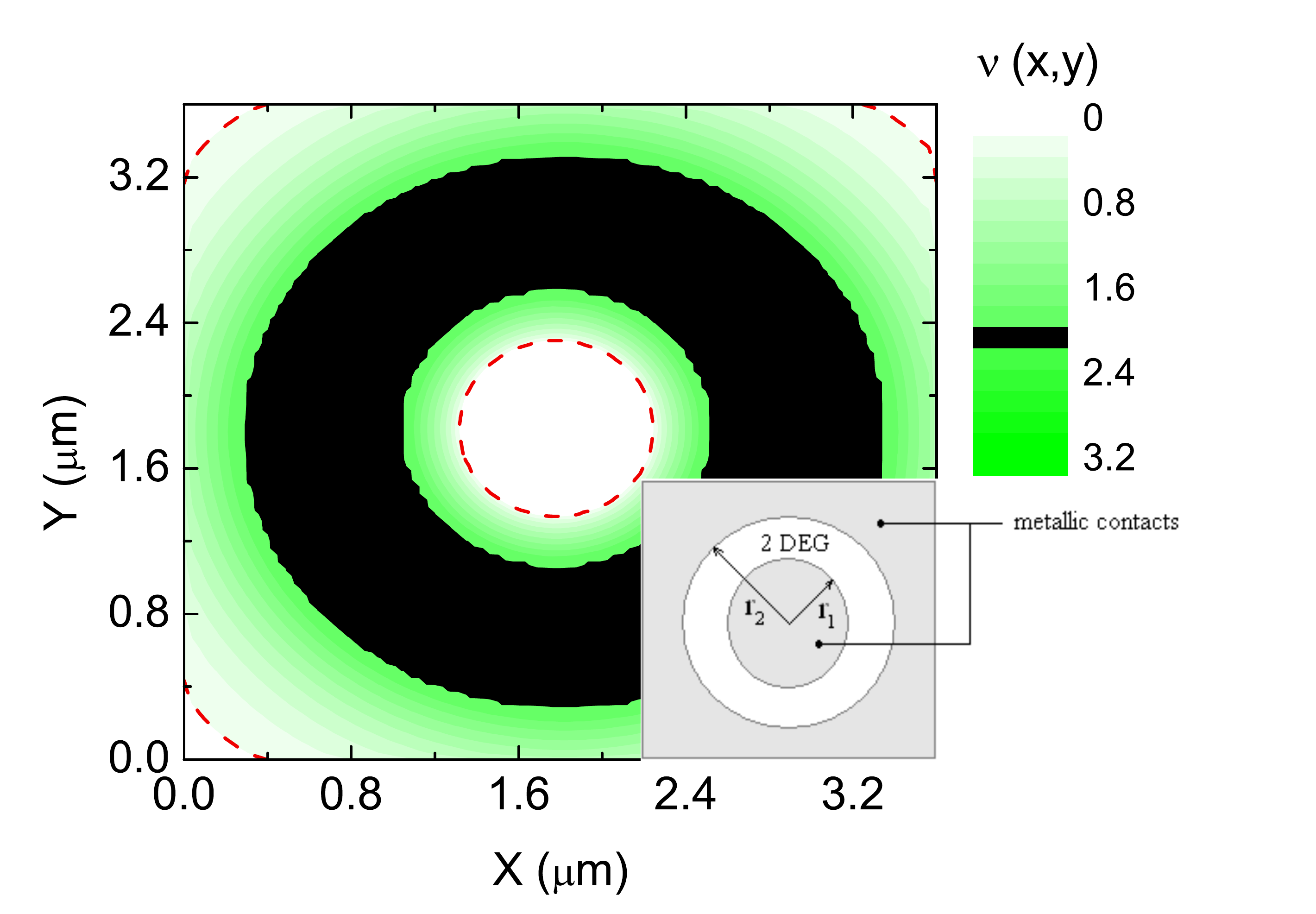}
\caption{\label{corbibulk}(Color online) Filling factor distribution of the disorder-free Corbino disc under $B=6.5$ T and $T=3$ K. The bulk of the sample is entirely incompressible, whereas Halperin-like compressible edge-states reside next to both inner and outer contacts. However, due to scattering within these regions quantization is not possible}}
\end{figure}

Here, we use the same self-consistent scheme introduced earlier. However, while considering the Corbino geometry, we do not impose an external non-equilibrium current via source and drain contacts. Instead, the inner and outer contacts are kept at a constant voltage, and only the external $B$ field is varied to observe the changes of the incompressible strips as a function of position. The Corbino disk that we study is an annular region of
a conductor (or an annular 2DES) surrounding a metallic
contact and surrounded in turn by a second metallic contact. The arrangements of the
contacts and the sample geometry are shown in the inset of Fig.~\ref{corbibulk}.
The circular contacts are designed with an inner radius of
$r_{1}=500$ $\rm nm$ and an outer radius of $r_{2}=1000$ $\rm nm$.
The dimensions of the outer metal contact are $3.6$ $\rm\mu
m\times3.6$ $\rm\mu m$. The confined 2DEG is
created by applying -5 V to the metallic contacts. Note that, due to the pinning of the Fermi energy at the surface (-.075 V), the gate's effective potential relative to the 2DES is $\lesssim- 4.25$ V.

\begin{figure}[t!]
{\centering
\includegraphics[width=1\linewidth]{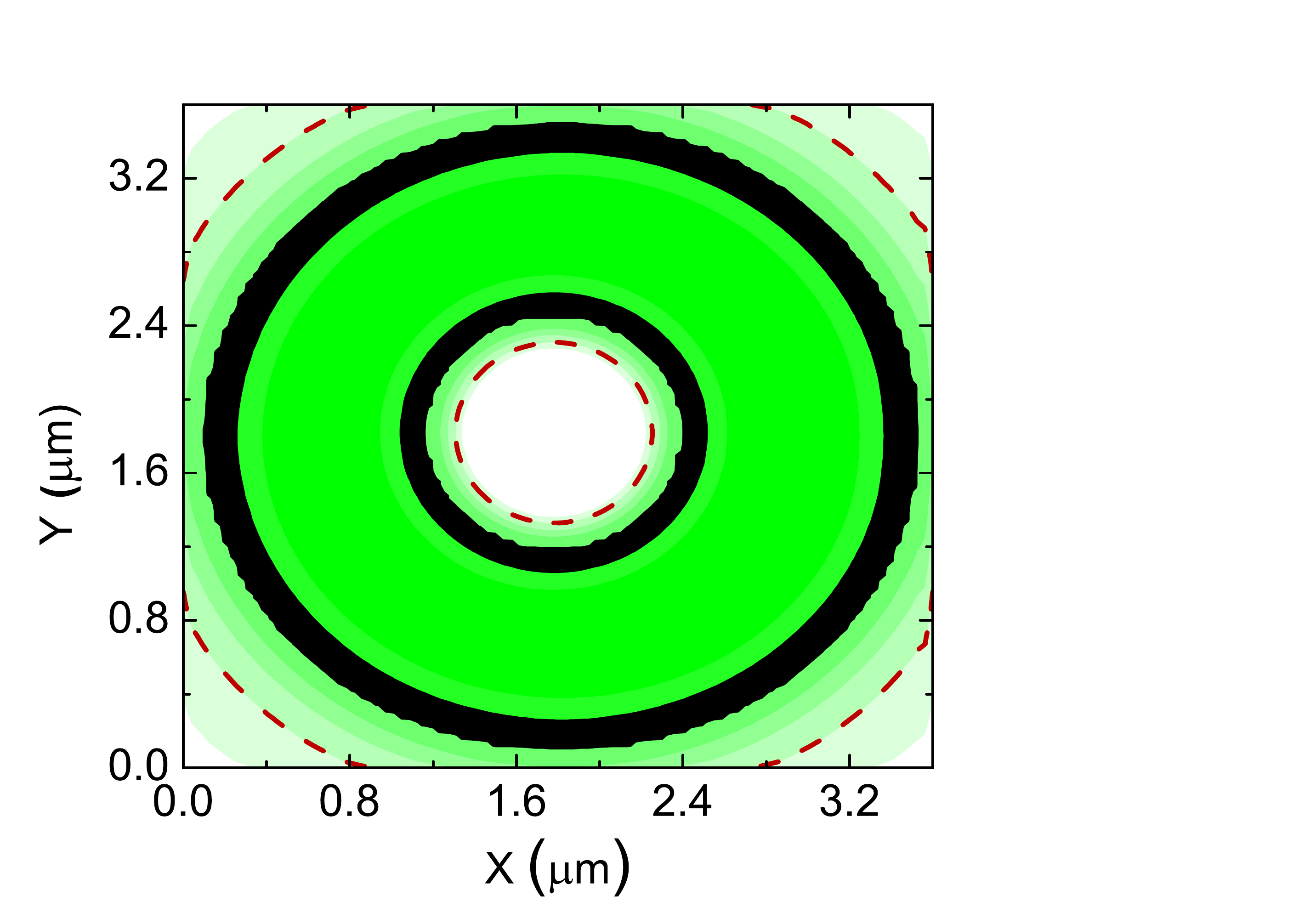}\caption{\label{corbiISs}(Coloe online) Same quantity as shown in Fig.\ref{corbibulk} at a higher magnetic field, $B=5$ T and same temperature. The incompressible edge-states are paralel to contacts, however, this situation can not be interpreted as the "absence of edge-states" (see the related text).}}
\end{figure}

Given the above structural conditions, we obtain the electron density (thereby the incompressible regions) and the electrostatic potential distributions, self-consistently described in the previous section. If the Fermi level is pinned to one of the Landau levels, then the system can be described by a metallic region with a finite density of states leading to relatively good screening. Hence, the electron density distribution presents a gradual change depicted by the colour gradient in Fig.~\ref{corbibulk} and Fig.~\ref{corbiISs}. The electron density effectively vanishes beneath the metallic contacts (white areas), bordered by broken (red) lines. On the contrary, if the Fermi energy lies within the Landau gap, the system is incompressible (also locally) due to the lack of states at the Fermi energy, and the density distribution remains constant throughout the incompressible region. The incompressible regions are depicted by dark (black) areas, where the electrostatic potential varies due to poor screening. The formation of incompressible regions is shown to be sensitive to temperature
and the strength of the magnetic field, i.e. if $kT/\hbar\omega_c\gtrsim0.05$, the incompressible region is smeared out.~\cite{Oh97:13519,siddiki2004,Bilayersiddiki06:} In this work, we will consider only the cases where the electron temperature is sufficiently low that incompressible regions are well developed.

\underline{Corbino geometry without disorder}

To investigate the topological aspects of a bulk insulator, we calculated the electron density distribution at relatively high magnetic fields, such that the bulk of the system becomes incompressible. Next, we show the local filling factor (which is nothing but a $B$ field normalized electron density distribution, namely $\nu(x,y)=2\pi\ell^2n_{\mathrm{el}}(x,y)$) profile at a characteristic magnetic field strength. Our calculations show that fixing $B$ to 6.5 T and decreasing the temperature down to 3 K extends the incompressible states such that the entire region of the disc proceeds bulk (Fig.~\ref{corbibulk}). Besides the compressible regions in front of the contacts, which is consistent with the scanning force microscopy experiments,~\cite{Ahlswede02:165,Dahlem10:121305} the electron density with an entire incompressible bulk corresponds to the Laughlin's bulk picture: To transfer a single charge from the inner contact to outer contact (or vice versa) adiabatically is protected by the ``topological insulator", and the Hall conductance is quantized due to phase constriction. The compressible edge regions near the contacts essentially do not destroy (even perturb) the phase argument. The influence of compressible strips on the observed quantities, such as conductance, is limited to measured contact resistance. The capacitive contribution to the conductance can be eliminated by calibrating the offset, utilizing a lock-in amplifier that can discriminate between the resistive and capacitive counterparts. It is important to note that the phase coherence length is much larger than the width of the compressible region, given the fact that the sample is high quality. Sufficiently small changes in the magnetic field strength would only enlarge (in the low field direction) or shrink (in the high field direction) the width of the incompressible bulk region. Hence, one can obtain a quantized Hall plateau of finite width. However, this is limited because this incompressible bulk could split into stripes residing at the opposite edges.~\cite{Oh97:13519} Our next step is to investigate such a situation.

We obtained circular incompressible stripes, in which the local filling factor assumes $\nu(x,y)=2$, near the inner and the outer annulus at the same temperature when the magnetic field is set to 5 T,
as shown in Fig.~\ref{corbiISs}.
The Corbino geometry, under the above conditions includes edge states
which remind Halperin's edge picture but are already different from his compressible
edge states. The bulk is compressible; however, the incompressible stripes decouple the inner and outer contacts. Therefore, even without the disorder, it is impossible to transfer an electron from the outer contact to the inner contact adiabatically. Namely the conductivity in the Hall direction (i.e. radial) is quantized due to incompressible strips. In other words, the incompressible (edge) stripes prevent scattering between the two contacts, even though the bulk is compressible. Having a compressible bulk is quite interesting since, if two contacts are embedded between the incompressible stripes (to visualize this configuration cf. Fig.~\ref{fig:hall_bar} and for simplicity assume that there are no poor density regions surrounding them), it is possible to measure a finite resistance between these two contacts. We mention that the bulk compressibility, even in the plateau regime, is justified by the scanning force microscopy experiments.~\cite{Ahlswede02:165,Dahlem10:121305} However, observed bulk compressibility is in contrast to previous experiments by Tsui and co-workers~\cite{Zheng85:5506} and also with the bulk picture of the IQHE.~\cite{Kramer03:172}

In summary, there are two incompressible stripes at the edges that decouple the inner and outer contacts. However, the bulk between these two stripes remains compressible, allowing charge transport between additionally embedded contacts. The scanning force microscopy experiments show the compressibility of the bulk. In contrast, it is observed that the bulk is incompressible within the plateau regime by earlier transport experiments compatible with the bulk theories. In the next step, we will show that both experiments and theories are consistent by including disorder to our calculations. In addition, as a direct test of the bulk compressibility, we propose that if two inner contacts are implanted to the bulk of the Corbino disc, one should measure a finite resistance even within the plateau interval, in contrast to the Laughlin's bulk and Halperin's edge picture. If this is the case, the system is not in a topologically protected state since such a result would show that scattering is possible within the bulk of the system.

\underline{Corbino geometry with disorder}

The above situation is altered if one considers a sufficient amount of disorder. Since disorder localizes electronic states also in the bulk even when there are two circulating edge incompressible strips. In order to analyze the disorder effects on the
edge-dependent description of IQHE, we include charged impurities to the entire Corbino disc with a disorder density of $~\% 4$ corresponding to
$n_{I}=356$ single impurities. Hence, the total potential variation is at the order of $\% 10$ of the Fermi energy, which satisfies the limiting conditions of the TFPA. In a recent investigation by Gulebaglan and co-workers
clearly shows that the long-range potential fluctuations become more extensive than the unit cell size if one considers less than $\% 5$ disorder.~\cite{Gulebaglan12:1495} Therefore, we will consider
only a sufficiently high number of impurities in our calculations to simulate a highly disordered system. Note that the disorder is always included in our calculations by short-range fluctuations that define the conductivities.
\begin{figure}[t!]{\centering
\includegraphics[width=1.0\linewidth]{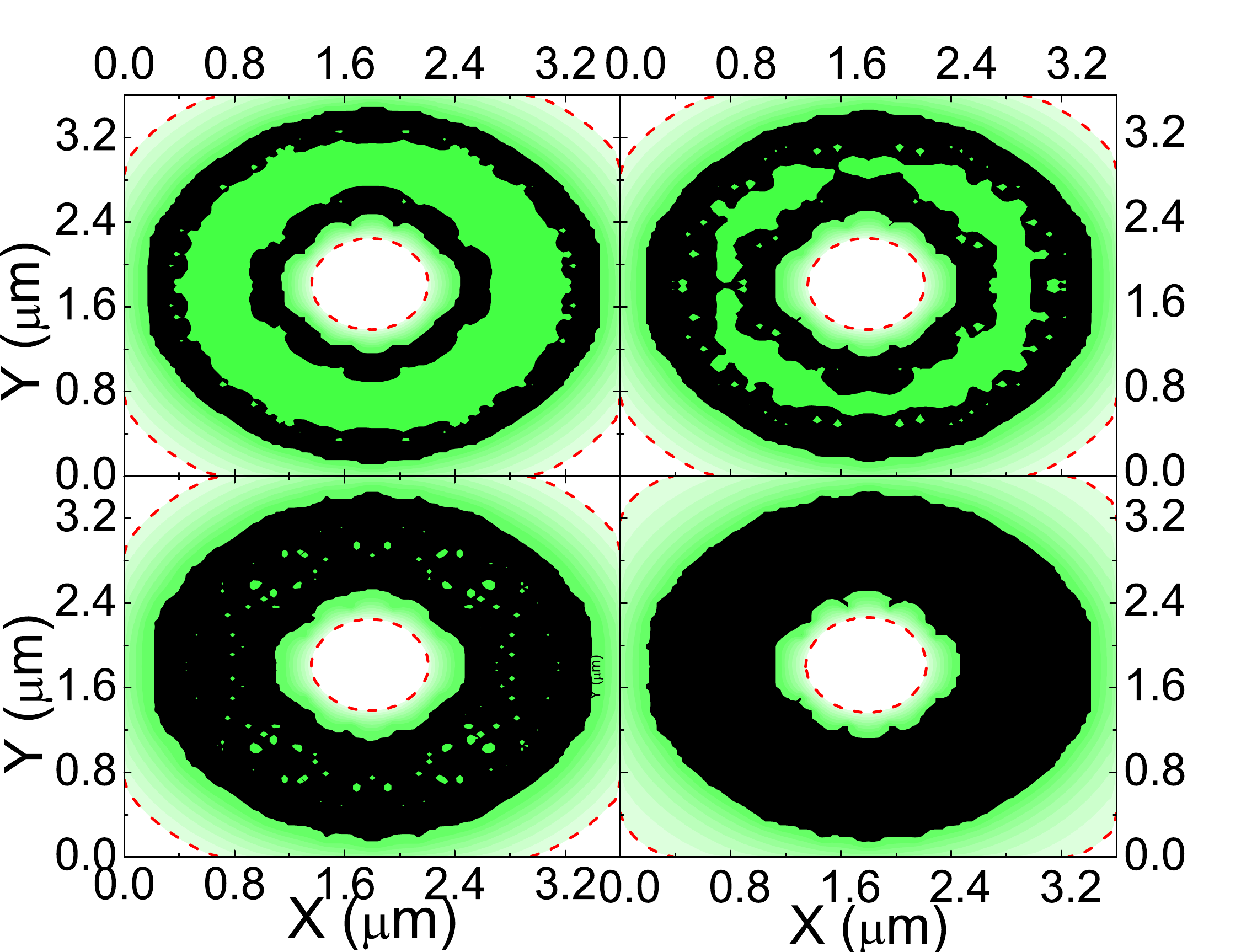}
\caption{\label{corbino_disorder}(Color online )Filling factor distribution of the Corbino disc with
the disorder at characteristic magnetic fields and temperatures (a) $B=6.2$ T, T=3 K, (b) $B=6.3$ T, $T=5$ K, 
(c) $B=6.4$ T, $T=7$ K and (d) $B=6.5$ T, $T=3$ K. We varied the temperature to clarify that, if the electrons thermal energy is much smaller than the cyclotron energy formation of incompressible regions remain approximately unaffected.}}
\end{figure}

We define the impurity potentials as Coulombic
\begin{equation}
V(r,z)=\frac{e^2/\overline{\kappa}}{\sqrt{r^2+z^2}},
\end{equation}
where $z$ denotes the spacing between 2DES and doping layer and
$r$ is a vector in $xy-$ plane. We solve the Poisson equation in 3D starting from the given material properties.
Then we construct a 3D lattice where the potential and the charge distributions are obtained
iteratively assuming open boundary conditions, {\emph{i.e.}} $V (x\rightarrow\pm\infty; y\rightarrow\pm\infty; z\rightarrow\pm\infty) = 0$.
We used the numerical simulation for the solution, which is based on a fourth-order
an algorithm is operating on a square grid in the $xy$ plane. This code is suitable for different boundary
conditions, applied successfully in previous studies.~\cite{Kilicoglu10:165308,Andreas03:potential,Sefa08:prb,Bilgecakyuz11:1514}

In particular, for the Corbino geometry using the periodic boundary conditions provide a closed-form, which
can be found in a well-known textbook,~\cite{Morse-Feshbach53:1240} on the other hand, for the Hall bar
geometry, we obtain the boundary conditions numerically.
We analyze the evolution of filling factor distribution in response to various magnetic fields
and temperatures, considering a ``dirty" sample. The inclusion of the disorder enlarges the incompressible edge strips as shown in Fig.~\ref{corbino_disorder}a and Fig.~\ref{corbino_disorder}b. At a higher magnetic field, the two edge stripes merge, and as a result, slight shifts in $B$ change the entire picture
from edge to bulk as shown in Fig.\ref{corbino_disorder}c and  Fig.\ref{corbino_disorder}d. By the inclusion of a disorder, it is observed that the bulk becomes incompressible for larger $B$ intervals; hence the bulk and the edge pictures merge and wider quantized Hall plateaus are observed by the experiments. If one implants two inner contacts similar to Hall bar geometry, no finite current can be measured between these contacts. Due to the localized states at the bulk. Hence, this situation coincides with the early Hall potential probe experiments~\cite{Zheng85:5506} and also nicely explains the current distribution at low-mobility large samples. Consequently, the topologically protected state is recovered due to disorder-induced localized states at the bulk. However, notice that in this case, one cannot talk about the compressible edge states of Halperin, since in this system, the edges are not defined by infinite potential walls. In contrast, the potential at the edges varies smoothly within quantum mechanical lengths.

As a final remark, we also note that performing similar calculations considering a Hall bar with high disorder yields analogous stripe enlargement with the Corbino disc. However, the picture does not change qualitatively. Therefore, we conclude that the edge and bulk insulators of the IQHE merge at highly disordered samples; hence the system's geometry becomes unimportant.

\section{Conclusion}

In this work, we investigated the influence of geometry on the so-called topological insulators of the quantized Hall effect, performing self-consistent numerical calculations. We first considered a Hall bar geometry where two independent inner contacts are embedded in the bulk, changing the genus number in real space. We argued that if the incompressible stripes reside at the edges connecting the injection contacts, one can drive a finite current between inner contacts without destroying the quantization of the Hall resistance. On the other hand, in the incompressible bulk case, we claimed that it is challenging to inject current between the inner contacts since the system's impedance would be relatively large.

In a further investigation, a long-range disorder fluctuation free Corbino disc at high magnetic fields showed that the bulk of the system becomes incompressible due to the Landau gap. Even solely the direct Coulomb interaction is taken into account. This situation coincides with the bulk picture of the IQHE. At lower $B$ fields, we observed two circular edge incompressible strips are formed, one encircling the inner and the other encircled by the outer contact decoupling them. Hence backscattering is suppressed by the incompressible edge strips and quantization of the Hall conductance reads. This case is somewhat similar to Halperin's edge-state picture, and however, in our calculation, the bulk is found to be compressible. Once the long-range fluctuations, i.e. high disorder, is considered, both cases merge, and the entire system becomes incompressible. The disorder broadens the incompressible regions, and the bulk picture is ascendant in both systems independent of the geometry. 

Note that, on the one hand, the incompressible edge strips obtained in the Hall bar sample are parallel to the straight line connecting the source and drain. On the other hand, the circular edge states in Corbino geometry cut the corresponding line perpendicularly. We conclude that the edge and bulk states are topologically related in momentum space. However, the topological characters of the Hall bar and Corbino disc systems are substantially different in real space. Furthermore, these two geometries coincide only if periodic boundary conditions are imposed, not describing a non-equilibrium dissipative Hall bar as measured in experiments.

In light of our calculations, we propose an experiment similar to Siddiki and co-workers, where at least two inner contacts are present at the bulk of the Hall bar geometry and measure the Hall (or longitudinal) resistance and capacitance separately. In already performed experiments, a substantially high potential difference is measured without influencing the quantization of the Hall plateau. Our proposed experiments expect to observe a resistance and capacitance quantization between inner contacts separately. Moreover, at the entrance and exits of each plateau, we expect to observe a non-linear resistive behaviour due to the transition from classical scattering based (Drude-like) transport to scattering free quantum tunnelling based (Landauer-like) transport.

In the second set of experiments, we suggest considering a relatively narrow Corbino disc ($r<20$ $\mu$m) to be defined on high mobility ($> 2\times10^6$ V.cm$^{2}$/s) wafers, similar in geometry with Tsui and co-corkers, however, higher in mobility and more petite in dimensions. In the case of edge incompressible strips, we expect to observe that half of the Hall voltage drops at one edge and the other half at the opposite edge, which the help of inner contacts can measure. In addition, if a finite current is imposed between longitudinally aligned inner contacts, we claim that one can measure a finite and quantized resistance between these contacts instead of substantial impedance.

The above-proposed experiments, we think, will clarify the bulk incompressibility of quantized Hall effect in real space by direct transport measurements and shed light on the family tree of the Topological Insulators.

\section*{Acknowledgements}
A.S would like to thank S. Sırt, S. Ludwig and S. Kumar for their fruitful discussions concerning the experimental realization of the samples and Klaus von Klitzing for giving the initial idea of considering inner contacts at a Hall bar geometry under IQHE conditions. T\"UBİTAK (The scientific and technological research council of Turkey) funded this project under 2219 program, and G\"OKABAD Research facilitated the computational power. A.S. is partially funded by ASDM financially.

\end{document}